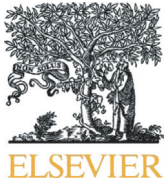
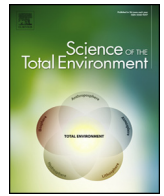

# Bioenergetics modelling to analyse and predict the joint effects of multiple stressors: Meta-analysis and model corroboration

Benoit Goussen [a,b,\*], Cecilie Rendal [b], David Sheffield [b], Emma Butler [b], Oliver R. Price [b,1], Roman Ashauer [a,2]

[a] *Environment Department, University of York, Heslington, York YO10 5DD, UK*
[b] *Safety and Environmental Assurance Centre, Colworth Science Park, Unilever, Sharnbrook, Bedfordshire, UK*

## HIGHLIGHTS

- Joint effects of natural and chemical stressors impact many ecological applications.
- Prospective ecological risk assessment requires an understanding of these effects.
- We lack full understanding of how organisms react to a combination of stressors.
- We used a Dynamic Energy Budget model to predict the effect of multiple stressors.
- We successfully showed the plausibility of our approach and validated our model.

## GRAPHICAL ABSTRACT

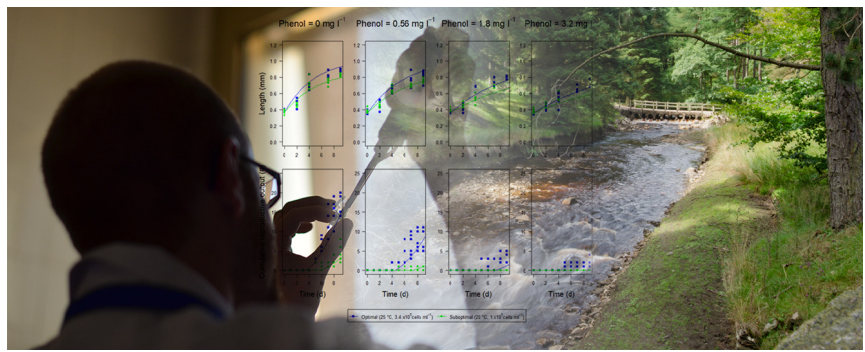

## ABSTRACT

Understanding the consequences of the combined effects of multiple stressors—including stress from man-made chemicals—is important for conservation management, the ecological risk assessment of chemicals, and many other ecological applications. Our current ability to predict and analyse the joint effects of multiple stressors is insufficient to make the prospective risk assessment of chemicals more ecologically relevant because we lack a full understanding of how organisms respond to stress factors alone and in combination. Here, we describe a Dynamic Energy Budget (DEB) based bioenergetics model that predicts the potential effects of single or multiple natural and chemical stressors on life history traits. We demonstrate the plausibility of the model using a meta-analysis of 128 existing studies on freshwater invertebrates. We then validate our model by comparing its predictions for a combination of three stressors (i.e. chemical, temperature, and food availability) with new, independent experimental data on life history traits in the daphnid *Ceriodaphnia dubia*. We found that the model predictions are in agreement with observed growth curves and reproductive traits. To the best of our knowledge, this is the first time that the combined effects of three stress factors on life history traits observed in laboratory studies have been predicted successfully in invertebrates. We suggest that a re-analysis of existing studies on multiple stressors within the modelling framework outlined here will provide a robust null model for identifying stressor interactions, and expect that a better understanding of the underlying mechanisms will arise from these new analyses. Bioenergetics modelling could be applied more broadly to support environmental management decision making.

\* Corresponding author at: ibacon GmbH, Arheilger Weg 17, 64380 Roßdorf, Germany.
*E-mail address:* benoit.goussen@ibacon.com (B. Goussen).
[1] Current address: RB, Dansom Lane, Hull, HU8 7DS, United Kingdom.
[2] Current address: Syngenta Crop Protection AG, Basel, Switzerland.

## 1. Introduction

Multiple stressors drive environmental change on a global scale, including the effects of the loss of biodiversity on ecosystem functioning (Vorosmarty et al., 2010; Rockstrom et al., 2009; Baert et al., 2016). Species are exposed to a multitude of natural and anthropogenic stressors, including man-made chemicals (Schwarzenbach, 2006; Schäfer et al., 2016). When factors in the ambient environment of a given organism fall outside their specific tolerance range, they become stressors. Consequently, organisms may alter their life history traits in response to these natural environmental stressors (e.g. low dissolved oxygen concentrations, food limitations, non-optimal temperatures, predation threats, or parasitism) (Boersma and Vijverberg, 1995; Hall, 1964; Homer and Waller, 1983; Seidl et al., 2005a; Connolly et al., 2004; Filho et al., 2011), including potentiation of effects or physiological mitigation (Hanazato, 1996; Penttinen and Holopainen, 1995; Seidl et al., 2005b; Coors and De Meester, 2008). This change in response to natural stressors has the potential to affect how organisms respond to chemical toxicants (Holmstrup et al., 2010; Laskowski et al., 2010).

Understanding the joint action of stressors can be a help in unravelling the combined adverse effects of multiple stressors that affect many ecosystems, and further, can be a contributing factor in understanding the qualities that new chemicals should have to exhibit high functional performance but limited effects on non-target environmental species (Schäfer and Piggott, 2018).

Prospective risk assessments of chemicals are widely used as a tool to help prevent potential risks to ecosystems and to develop more sustainable products. Current environmental risk assessment approaches are typically aimed at assessing single chemicals, and although mechanistic understanding of the biological and chemical processes that govern adverse effects is often incorporated into assessments, most assessments still rely on single chemical and single species interactions. A key challenge is to increase the ecological relevance of the prospective risk assessment of chemicals by tackling the effects of multiple stressors on multiple species (Scientific Committee on Health and Environmental Risks SCHER, 2013; van den Brink, 2008; Relyea and Hoverman, 2006; Rohr et al., 2006). Specifically, our ability to predict the impact of multiple simultaneous environmental and chemical stressors is hampered by a lack of understanding of how these factors interact to alter life history traits (Holmstrup et al., 2010; Galic et al., 2018; Crain et al., 2008).

The current prospective environmental risk assessments (ERAs) of man-made chemicals has been developed to be protective rather than predictive and is widely recognised to lack ecological realism (Forbes and Calow, 2013; Goussen et al., 2016; Forbes et al., 2009; Ashauer et al., 2011). This is in part rooted in the way standard endpoint data are derived. In ERAs, organisms are tested under laboratory experiments at strictly-controlled—and often optimal—conditions (e.g. ad libitum food intake, optimal temperature, high levels of dissolved oxygen) in order to reduce the effects of confounding factors and isolate the effect of the chemical. Even when studies have assessed the combined effects of natural and chemical stressors (Holmstrup et al., 2010; Chandini, 1988; Hamda et al., 2014; Cedergreen et al., 2016; De Coninck et al., 2013; Pieters and Liess, 2006) they did not result in a general mechanistic theory or predictive models. This is the consequence of the fact that ERAs must simplify some of the environmental complexity to be applied on a daily basis. Consequently, only limited efforts have been made to develop tools like mechanistic and predictive models accounting for multiple stressors. However, we need such tools, because accounting for the effects of multiple stressors—both chemical and natural—and allowing extrapolating to untested species or combinations of stressors and stress intensities would improve the ecological relevance of ERAs.

When the observations of multiple-stressor experiments deviate from the joint effect predicted by a given null model (i.e., the reference model), researchers typically infer that an interaction—specifically synergism or antagonism—must have occurred (see Holmstrup et al., 2010; Laskowski et al., 2010; Crain et al., 2008 for meta-analyses). In these cases, the terms synergy and antagonism are used as black box terms to describe an unknown effect that has caused a greater or smaller response than expected. However, in many cases, the underlying null models do not reflect the true system, even in the absence of interactions because they are poorly defined, and lack a mechanistic basis (Schäfer and Piggott, 2018). Thus, it is possible that studies reporting synergy and antagonism in the face of the multiple stressors could be refined to confirm and better understand the reported effects.

Furthermore, while the null models that are typically applied at organism level may consider the statistical distribution of sensitivities and the observed stressor-effect relationships, they rarely model the underlying processes (e.g. physiology), even though the central role of organism level-responses in stress ecology has long been understood (Maltby, 1999). Thus, we argue that a step change is needed in the analysis and prediction of the effects of multiple stressors. Here we propose a generalised, mechanistic framework for predicting the joint effects of stressors on the life history traits of a given individual organism. By modelling stressor-induced changes in energy allocation, we propose to predict the impact of a multiple-stress environment on an organism's growth and reproduction, and therefore on population dynamics.

Organismal development is a thermodynamic process. Organisms require energy to maintain their basal metabolism, move, create body mass, reach sexual maturation, and reproduce (Dajoz, 2006). Bioenergetic models, such as the Dynamic Energy Budget (DEB) model (Kooijman, 2010) are coherent models of organism physiology, and integrate various physiological processes using a consistent, quantitative, and rigorous framework. DEB models describe how energy is assimilated from food and used for the organism's maintenance, growth, maturation, and reproduction following the $\kappa$ rule (van der Meer, 2006), where a fraction of the energy ($\kappa$) is allocated to maintaining homeostasis (i.e. somatic maintenance) and growth processes. The remaining energy $(1-\kappa)$ is allocated to processes associated with maturation and reproduction. DEB models can pinpoint which life history processes are perturbed by anthropogenic (e.g. chemical) or natural stressors (Freitas et al., 2011; Goussen et al., 2015; Jager and Zimmer, 2012; Ashauer and Jager, 2018), or combinations of these stressors (Pieters et al., 2006; Jager et al., 2010), where the resulting impacts on growth, maturation, and reproduction are easily translated into changes at the population level (Beaudouin et al., 2015). In that way, the results from highly standardised laboratory toxicity tests can be extrapolated to ecologically relevant, field conditions (David et al., 2018).

We can now see that bioenergetics modelling with DEB models could offer a coherent framework to understand and predict effects of multiple stressors acting individually or jointly on organisms and so increase the ecological relevance of ERAs. What we do not yet know is how well this idea works. Therefore, the aims of the present study were: i) to develop a bioenergetics-based modelling framework to predict the effects of single or multiple natural and chemical stressors on life history traits and use the extracted and normalised data from previously published studies to assess the model's plausibility, and ii) to corroborate the model by comparing its predictions with novel experimental data.

## 2. Materials and methods

### 2.1. Meta-analysis

As freshwater invertebrates have a long history as model test organisms, the relatively large amount of data available in the published literature allows us to interrogate of the role that the interaction of environmental and chemical stressors plays in altering the life history traits. To this end we performed a meta-analysis of the scientific literature (see details in Supporting Information (SI), Table S1).

For each study collated from the literature, a control (or optimum) condition was chosen, which correspond to the condition with the strongest life-history trait values (e.g. largest body size, most reproductive output). Typically, these are conditions with ad libitum food availability, high oxygen levels, absence of extremes of temperature, and no addition of toxic compounds. Data extracted from each of the studies were scaled by the optimal value (e.g. largest body size or maximum reproductive output) for the corresponding study to facilitate comparison.

### 2.2. DEB simulations

#### 2.2.1. Bioenergetics modelling: DEBkiss model for D. magna

To assess the ability of DEB models to predict effects of multiple stressors, we used a DEB-based model for *Daphnia magna* to simulate how single stressors and multi-stressor combinations would affect the life-history traits of organisms. We then compared these simulations to the meta-analysis results. We simulated the effect of single stressors using the DEBkiss model, which is a simplified application of DEB theory (Jager et al., 2013) that has been successfully applied on a wide range of organisms (see growing list on http://www.debtox.info/debkiss_appl.html). The simulations were performed on *D. magna* using a model previously established parameter values (Jager, 2017) (see SI Sections 2.1.4 and following).

For each individual stressor of interest, the relevant DEBkiss model parameter values (e.g. parameters linked to food for the food level stressor) and variables were varied in a range that corresponded to the range of the data collected in the meta-analysis. Simulations were run for 30 days, which was sufficient time to reach the maximum (asymptotic) body length under optimal conditions. The outputs analysed were (i) the body length at the end of the simulation under a wide range of sub-optimal conditions (i.e. stress) divided by the simulated maximal body length the organism can reach under optimal conditions (i.e. no stress) and (ii) the cumulative reproduction (i.e. offspring per female) at the end of the simulation under a wide range of sub-optimal conditions (i.e. stress) divided by maximal cumulative reproduction (i.e. offspring per female under no stress conditions). This procedure allowed us to obtain normalised endpoints that can be compared across different studies of different species in the meta-analysis. The simulation results and the data from the meta-analysis are then plotted as overlays across the range of the individual stressor (Fig. 1).

The effect of the combination of two stressors was also simulated with DEBkiss using the same strategy as for the effect of each individual stressor of interest and parameters for *D. magna*, with the exception that Monte Carlo simulations were performed using the ranges of the model parameters or input variables (e.g., food availability, temperature, presence of predator kairomones, cadmium levels) that corresponded to those collected for the two stressors of interest. The outputs of these simulations were used to plot raster and contour plot overlays (Fig. 2).

### 2.3. Case study with Ceriodaphnia dubia

#### 2.3.1. Experiments

To test and corroborate our multiple-stressor response model, an extensive set of laboratory experiments were performed using the cladoceran *Ceriodaphnia dubia*. A brief summary of the design is given here; detailed information are available in the Supporting Information. Specifically, we developed a high-throughput experimental method for measurements of growth, reproduction, and mortality in *Ceriodaphnia dubia*. The short life cycle of *C. dubia* allows for chronic toxicity assays to be performed in 7–10 days (compared to *Daphnia magna* at 21 days). To assess the effects of food availability and temperature on growth and reproduction, we used a factorial study design in which *C. dubia* were exposed to one of four different temperatures (15, 20, 25, and 30 °C) and one of three different concentrations of *Chlorella vulgaris* ($0.5 \times 10^5$, $4 \times 10^5$, and $15 \times 10^5$ cells mL$^{-1}$). In addition a commercial food supplement was provided to all batches (GP 5-50), which contains marine fish, krill (23%), fish roe, soy lecithin, yeast autolysate, microalgae, fish gelatine, squid meal, hydrogenated vegetable fat, vitamins and minerals, antioxidants. In spite of the conventional duration of 7–10 days for the *C. dubia* toxicity test, our experiment was conducted for 21 days, to ensure that asymptotic growth and reproduction were reached for organisms cultured under optimal conditions.

Two scenarios were investigated in combination with the effects of phenol on growth and reproduction: (A) one in which organisms were exposed to phenol under conditions of optimal temperature and food availability and (B) one in which they were exposed to phenol under sub-optimal temperature and reduced food availability. In scenario A, organisms were exposed to phenol (nominal concentrations of 0.0, 0.32, 0.56, 1.0, 1.8 and 3.2 mg L$^{-1}$) under standard (optimal) culturing conditions (25 °C and a feed level of $3.4 \times 10^5$ cells mL$^{-1}$ *C. vulgaris* + GP 5-50). In scenario B, organisms were exposed to the same concentrations of phenol as Test A, but at a slightly lower temperature (20 °C) and a significantly reduced concentration of *C. vulgaris* ($1 \times 10^5$ cells mL$^{-1}$ + GP 5-50).

#### 2.3.2. Bioenergetics modelling

A bioenergetic model based on the DEB theory was calibrated and used to perform independent simulations. A full description is available in the Supporting Information Sections 2.3 and 2.4. Briefly, a DEBkiss model was calibrated in two separate steps. In the first one, the calibration was performed on growth and reproduction data obtained from an experiment with food and temperature stress but without chemical stress. In the second step, the toxicity parameters were calibrated on data from an experiment with chemical stress without food or temperature stress. Without further calibration, the DEBkiss model was then used to predict the growth and reproduction over time for *C. dubia* exposed to either optimal food and temperature conditions and phenol (i.e. scenario A) or to a non-optimal temperature, food level, and phenol concentration (scenario B).

## 3. Results and discussion

### 3.1. Meta-analysis results for single stressors

#### 3.1.1. Comparing empirical data from different species with a model of daphnids

The effects of temperature, food level and oxygen on growth and reproduction in different freshwater invertebrate species are too sparse and too variable to extract any information on the differences between the species (Fig. 1). Limited information can be extracted from the relationship between the temperature and growth and reproduction (Fig. 1A, B) because the data points are all in the 70% to 100% range. The relationship is more defined for food levels (Fig. 1C, D) and oxygen (Fig. 1E, F), where a pattern can be seen in the empirical data. Overlaid with this empirical data is the bioenergetics (DEB) model prediction for effects of temperature, food level and oxygen on growth and reproduction in *D. magna*. The model line goes through the data in all panels of Fig. 1, which tells us that the model for *D. magna* - scaled in the same way as the experimental data in order to facilitate comparison - is consistent with empirical data for a total of 15 different species, although the empirical data is very patchy and variable. Thus, to improve theory and model, we need more data to better cover the whole range of stressor values, especially for temperature, and we need less variable data.

#### 3.1.2. Temperature

The influence of temperature on metabolic rate has been well documented (Kooijman, 2010). Our meta-analysis included 10 studies with data on the effects of temperature stress on the maximal length and cumulative reproduction for 7 species (Fig. 1A, B) and revealed that data were primarily collected well within the temperature tolerance ranges of the species studied.

Few data points fall outside the boundaries of the tolerance ranges. This is most likely because researches avoid the experimental complications expected at temperature extremes. At low temperatures long experimentation times are required, and at high temperatures, high mortality rates are expected.

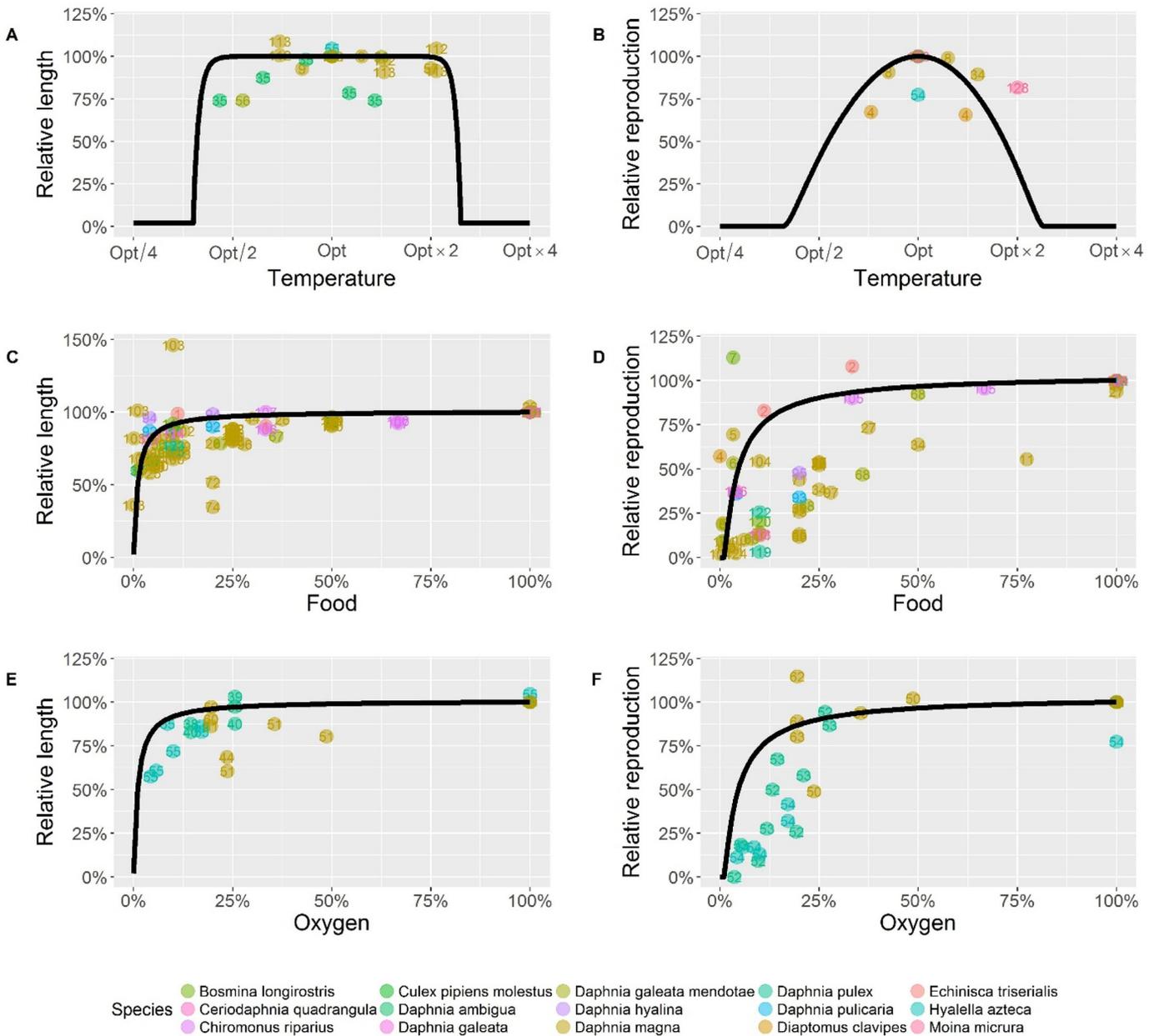

**Fig. 1.** Effect of environmental stressors on body length and reproduction of freshwater invertebrates. Data from the meta-analysis are presented by points. Black lines represent Dynamic Energy Budget (DEB) simulations of *Daphnia magna* responses. Body length and reproduction data are presented relative to the values of the optimum condition achieved under non-stressed conditions (see meta-analysis methods for definition).

Our analysis also shows that the temperature relationship used in our model for *D. magna* is consistent with the data on temperature and life-history traits obtained for 7 species. In the DEB model simulation of reproduction is compromised by even a small deviation from the optimal temperature (Fig. 1B). In contrast to reproduction, growth is maintained over a wider temperature range, followed by a steep reduction at the boundary of temperature tolerance (Fig. 1A). Additional experiments should consider adding data points that are close to the maximum and minimum temperature tolerance values of each species because these data are lacking (Fig. 1A, B). This would facilitate better model calibration and extrapolation to a wider range of temperatures.

### 3.1.3. Food

The meta-analysis of the literature yielded 179 data points from 13 species related to effect of food density on growth and reproduction. We found that a slight decrease in food density from the optimum food level induces only a small effect on maximal body length and cumulative reproduction (Fig. 1C, D). In contrast, low food density leads to a large impact on both growth and reproduction (Fig. 1C, D). The decline in relative reproduction is more pronounced and appears already at higher food levels than the decline in body length (Fig. 1C, D). That suggest that a reduction in food availability affects reproduction before it affects growth and that this effect is more pronounced with decreasing food levels. This pattern from our meta-analysis is a good match with the results we obtained using our DEB model for *D. magna*. Although the effects of food availability on growth and reproduction are generally well captured by the DEB model, the large degree of variability observed for reproduction data across food levels (Fig. 1D) means that little general knowledge can be gained here. This could imply that we currently do not know how reliable DEB model predictions at different food levels are or it could imply that life-history traits vary considerably in response to food levels below ad libitum. Optimal food density for a given species can be highly variable depending on the study. For instance, while *D. magna* has previously been considered to be fed ad libitum at food densities of 0.75 to 2 mg C L[1] (Heugens et al., 2006; Pieters et al., 2005), it has been suggested that some of these densities might actually be food-limiting (Augustine et al., 2011).

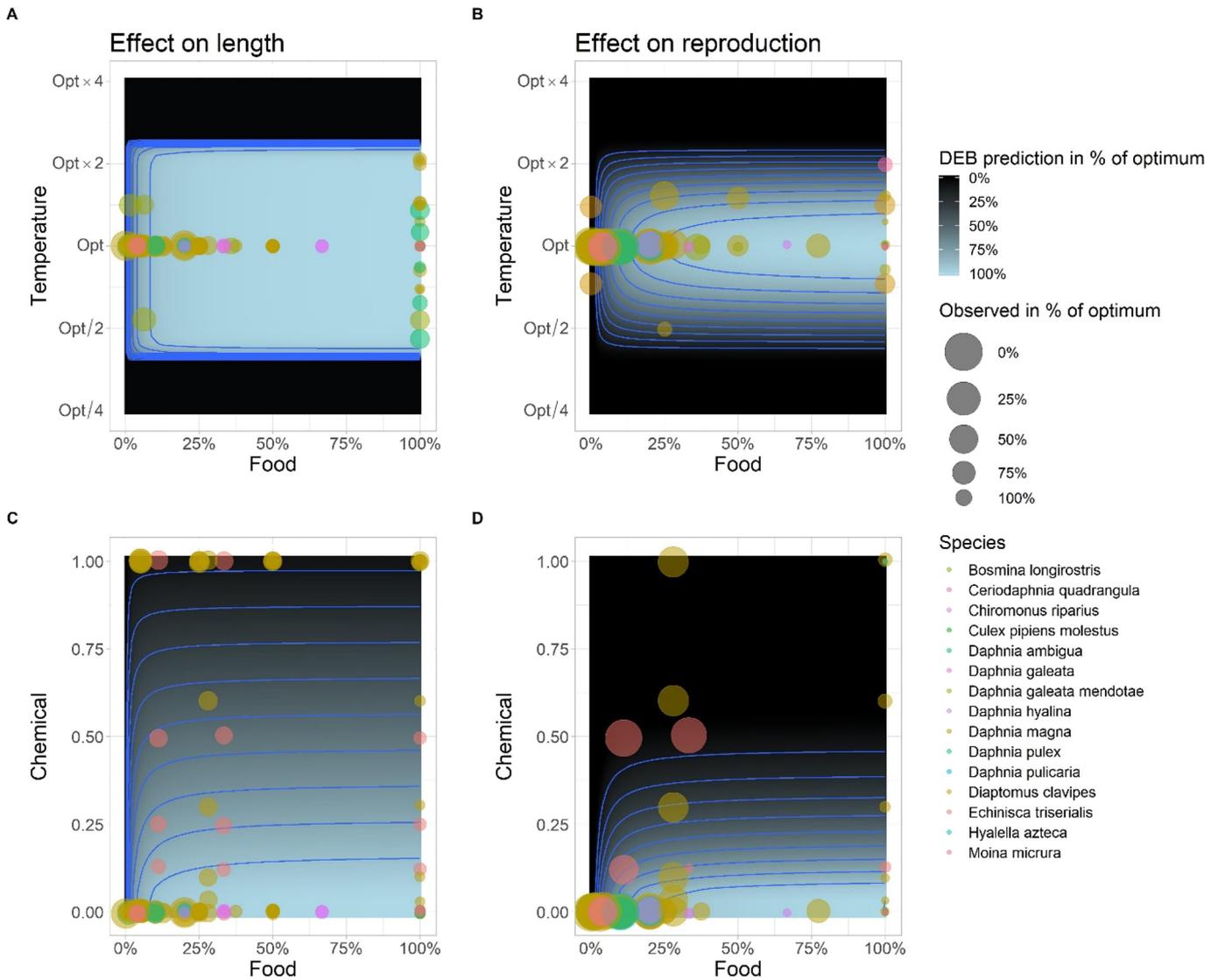

**Fig. 2.** Joint effects of environmental and/or chemical stressors on length and reproduction. Data from the meta-analysis are presented as points of increasing size (i.e. greater effects are represented by larger sizes). DEB simulations are represented by response surface and contour plots. Effects are presented as relative to the amount of growth (i.e. body length) or reproduction achieved under non-stressed environmental conditions. Stressor levels are represented as relative to an optimal condition (e.g. 0% represents an absence of food compared to the optimal condition, and 100% represents 100% of the optimal food availability). The chemical modelled for panels C and D is cadmium. See Supporting Information for raw data.

### 3.1.4. Dissolved oxygen

The meta-analysis shows that the three freshwater invertebrate species where data were available are tolerant to reductions in dissolved oxygen levels down to approximately 25% of the optimal concentration (Fig. 1E, F; where optimal is as defined for the meta-analysis, typically ranging between 7 and 9 mg L$^{-1}$), below which the impact of dissolved oxygen on both growth and reproduction increases rapidly. In other words, organisms can compensate for sub-optimal environmental conditions to a critical cut-off point, after which compensatory mechanisms become ineffective. Our modelling results are consistent with the picture from the meta-analysis (Fig. 1E, F), which confirms that our simple modelling approach accounts for the effect of low levels of dissolved oxygen on life history traits of freshwater invertebrates. However, it is important to note that the onset of the effects appears at slightly higher oxygen levels in the empirical data compared to the model. Future studies could explore if this mismatch could be eliminated by using the principle of the synthesising units (Kooijman, 2010) instead of the very simple modelling approach used here. The synthesising unit concept could play an important role in increasing the coherence of the model with the empirical data as it is a generalised unit that processes incoming substrate (e.g. food, compound) in order to yield one or more products (Muller et al., 2019).

### 3.2. Multiple stressors

#### 3.2.1. Combination of sub-optimal food levels and temperature

Food and temperature play a role in determining both the energy available to and the energy required by a given organism, via their effects on metabolic rates. They are thereby among the main drivers of development. The meta-analysis indicates that these two stressors jointly affect both growth and reproduction, although reproduction is impacted to a greater degree (Fig. 2). A limitation is that the meta-analysis data are quite sparse outside of or close to the temperature tolerance ranges of the studied organisms. Data are also scarce for intermediate food levels at non-optimal temperatures. According to DEB theory, although growth remains relatively unchanged within the temperature tolerance range, steep changes occur at the boundaries of this range. In contrast, the negative effects of low food availability appear only at very low food levels. Given the scarce and unevenly distributed meta-analysis data relating to body length (Fig. 2A), it is difficult to assess whether the pattern predicted by DEB is reasonable. The data on reproduction (Fig. 2B) at deviations from optimal conditions indicates a shallower gradient towards impacts compared to the data for growth (Fig. 2A), but the impact starts at smaller deviations from optimum conditions. While this pattern is well captured by the DEB model, only a few data points are available where the predicted gradient is the steepest. The DEB theory assumes that reduced food availability reduces the amount of energy available, while temperature acts on metabolic rate.

Both reproduction and growth appear to be impacted by the interaction between food and temperature. For reproduction the impact occurs at small deviations from optimum conditions with a shallow gradient (Fig. 2B). For growth the impact occurs at stronger deviations from the optimum but with a steep gradient (Fig. 2A).

### 3.2.2. Combination of sub-optimal food levels and a toxicant

Both food and toxic chemicals have the potential to impact the life history traits of organisms. The example presented here is based on the joint effect of cadmium and food availability on growth and reproduction (Fig. 2) (Table S1). A limitation of the empirical meta-analysis data is that data are scarce, especially data for high chemical concentrations and intermediate food levels. Our model simulates a limitation in the intake of available energy (caused by a reduction in food availability) coupled with the physiological mode of action of the chemical, which in this case, impacts the assimilation of energy from food (Margerit et al., 2016; Baillieul et al., 2005). Other physiological modes of action can be modelled in similar fashion. The meta-analysis results (empirical data and model) highlight the interactive effects of the two stressors on both growth and reproduction, and shows that the joint stressors appear to exert stronger effects on reproduction than on growth (i.e. body length). This trend is particularly evident (in both the empirical meta-analysis data and the model) as a result of increasing concentrations of the chemical in question. More specifically, while a slight increase in concentration appears to have almost no effect on growth, our model shows that it can reduce reproductive output by as much as 75%.

Assessing this type of combination of potential stressors is of great interest for environmental risk assessments, as these stressors can alter the effects of each individual stressor. Furthermore, modelling the interactions of these stressors allows for the extrapolation of chemical effects from an environment with a constant concentration of food (e.g. laboratory conditions) to an environment with fluctuations in food availability (e.g. field conditions).

### 3.3. Case study: Predicting combined effects of chemical and environmental stress

To better evaluate the ability of DEB theory to predict the combined effect of natural and chemical stressors on growth and reproduction, modelling should be performed in conjunction with physiological case studies. The following case study outlines laboratory studies using the daphnid *C. dubia* exposed to three stressors: temperature stress, food stress, and chemical (i.e. phenol) stress. An in-depth analysis of the case study experimental results can be found in the Supporting Information Section 4.

#### 3.3.1. Calibration of the effects of food stress and temperature

We first assessed the effects of combined food and temperature stress, by testing three food levels at four temperatures. Statistical analysis shows that temperature affected both growth and reproductive traits in *C. dubia*, and specifically impacted growth rate, final length, and cumulative reproduction (Fig. 3). Reproductive traits, in particular, proved to be sensitive to temperature and were also highly impacted by reducing food levels at all temperatures. As the experimental temperature increased, only the lowest food level resulted in additional impacts on growth and reproductive traits. At the optimal temperature (25 °C), growth was impacted in a dose-dependent manner based on food level, while reproductive traits were only impacted at the lowest food level. A similar pattern was found at the highest temperature tested (30 °C), although the average final length of *C. dubia* was smaller than that of organisms exposed to their optimal temperature. The observed effects of the joint stressors (food density and temperature) on growth and reproductive traits in *C. dubia* were well captured by the DEB model with a tendency to overpredict the effect on reproduction at higher temperature and lower food level as shown in Fig. 3. The temperature and feeding parameters obtained are presented in Table S4.

#### 3.3.2. Calibration of the chemical stress

Next, chemical (i.e. phenol) stress was introduced into the experiment. We designed an experiment using optimal food levels and temperature and exposure to a range of phenol concentrations to infer the toxicity related model parameter values for phenol. Exposure to phenol under otherwise optimal conditions (25 °C, $3.4 \times 10^5$ cells mL$^{-1}$ of *C. vulgaris*) resulted in a dose-dependent response on both growth and reproduction (Fig. S2). As was the case with food and temperature stress, the effects of phenol on *C. dubia* were well captured by the DEB model, with a no-effect concentration (NEC) of 0.142 mg L$^{-1}$, a concentration tolerance ($c_T$) of 1.524 mg L$^{-1}$, and a dominant rate constant ($k_e$) of 10 d$^{-1}$, which corresponds to an almost immediate internalisation of the compound in the organism (Fig. S2 and Table S3).

#### 3.3.3. Simulation of two scenarios

Based on the above simulations of experimental results, all DEBkiss model parameters (i.e. those related to basic organism physiology, temperature, food, and chemical toxicity) were fixed, and a DEBkiss simulation was used for two scenarios exposed to a range of phenol concentrations: (A) one based on the standard set of optimal conditions typically used in a laboratory (e.g. 25 °C, $3.4 \times 10^5$ cells mL$^{-1}$ of *C. vulgaris*), and (B) the other scenario with mild temperature and food stress (20 °C, $1 \times 10^5$ cells mL$^{-1}$ *C. vulgaris*), which exemplifies extrapolation to a more environmentally realistic situation. The results of this simulation were then compared with the experimental results. As growth and reproduction were already impacted by the non-optimal environmental conditions, the experimental results showed that phenol had a dose-dependent impact on both growth and reproduction, with a statistically greater impact under non-optimal conditions (see SI for statistical analysis). Likewise, these patterns were accurately predicted by the simulation (Fig. 4), thereby demonstrating the extrapolation from standard environmental conditions to non-standard conditions involving complex combinations of stressors.

#### 3.3.4. Effects on survival

It is important to note that survival was also impacted by both environmental conditions (food level and temperature) and phenol toxicity. *Ceriodaphnia dubia* exposed to varying levels of food densities and temperatures exhibited high mortality at low temperatures for all food densities as well as at the lowest feed density for all temperatures (Table S5). In addition, mortality at day 21 was greater than 80% and as high as 100% at the highest temperature. Phenol also induced mortality in a dose-dependent manner, where we observed a greater impact under non-optimal conditions (e.g. up to 100% mortality at >1.8 mg L$^{-1}$) than under optimal conditions (e.g., up to 66% mortality at 3.2 mg L$^{-1}$). In the current study, only the sublethal effects have been investigated. The survival effect could be analysed by adapting a model like the General Unified Threshold model of Survival (Jager et al., 2011; Jager and Ashauer, 2018) to account for a combination of environmental and chemical stressors.

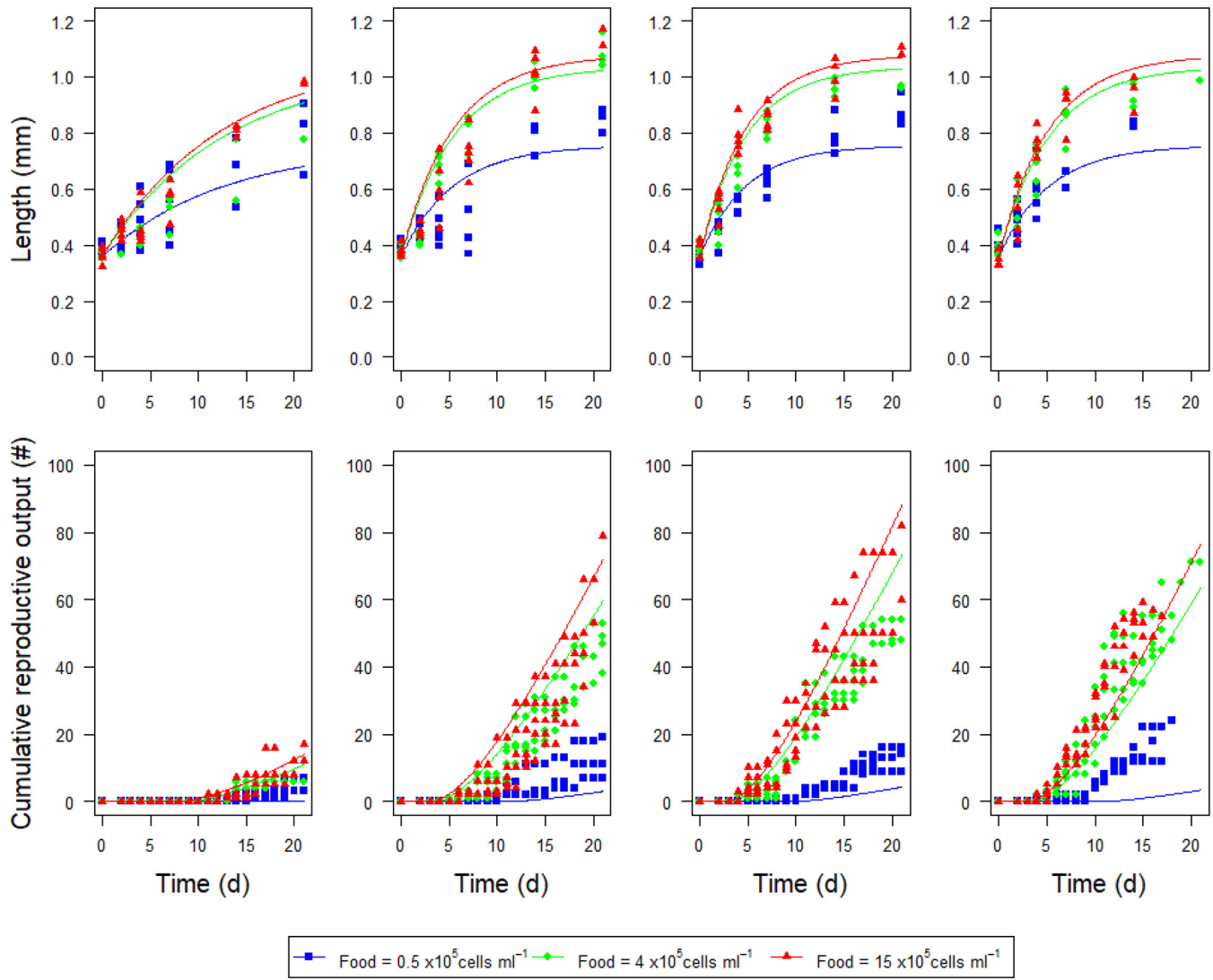

**Fig. 3.** The joint effects of temperature and food density on growth and reproductive parameters in *C. dubia*. The upper graphs represent growth (i.e. length) over time, and the lower graphs represent the cumulative reproductive outputs over time for four different experimental temperatures and three food densities ($0.5 \times 10^5$, $4 \times 10^5$, and $15 \times 10^5$ cells ml$^{-1}$). Our experimental data are represented by squares, triangles, and lozenge (one point per surviving organism at each time-point), while the dynamic energy budget model simulations (fitted) are represented by the lines.

### 3.4. Research needs

#### 3.4.1. Additional stressors and real-world applications

Accounting for the impact of other environmental variables that can become stressors—such as pH, nitrates, or nitrites—can be important in environmental risk assessments, particularly if one is interested in the life history traits of organisms in the "impact zone" (i.e. downstream) of wastewater discharge points (Finnegan et al., 2009). This is particularly true in many parts of the world where untreated wastewater is routinely discharged into surface water. Stressors such as pH or nitrate can severely alter the life history traits of organisms (Soucek and Dickinson, 2016; Rendal et al., 2012; Davis and Ozburn, 1969; Alibone and Fair, 1981) and could be included in bioenergetics models in the same manner as chemical toxicants. Furthermore, although life history traits can also be altered by biotic stressors that are widely encountered in the natural environment (e.g. pressures owing to competition, parasitism, or predation), these stressors are not usually accounted for in classical ERA frameworks, and the scientific understanding of how to quantitatively model their interactions with chemical stress is limited. In a comparable manner, the current understanding on how to quantitatively integrate environmental factors with high potential impact on the biodiversity such as habitat destruction or hydro-morphological alterations in a multiple-stressor environment requires further scientific development. Such factors would require models focusing on a higher level of organisation such as individual based models. The present study, although it focuses on individual level modelling, can serve as a guide to developing models for these additional stressors.

#### 3.4.2. Modelling and theory

Here, we show that the DEB model—which includes both environmental and chemical stressors—can accurately predict patterns in the effects that result not only from exposure to a single stressor (Fig. 1), but also a mixture of two (Figs. 2 & 3) or three (Fig. 4) stressors. Our analysis also highlighted gaps in the existing experimental data, particularly in the stressor ranges that result in steep model responses (e.g. borderline temperatures; Fig. 1A and B). Further work is required to assess the predictive capabilities of bioenergetics modelling for a wider range of stressor combinations, especially those with three or more stress factors. Mapping the relative importance of stressors in real environments would also inform risk management, conservation, and mitigation strategies.

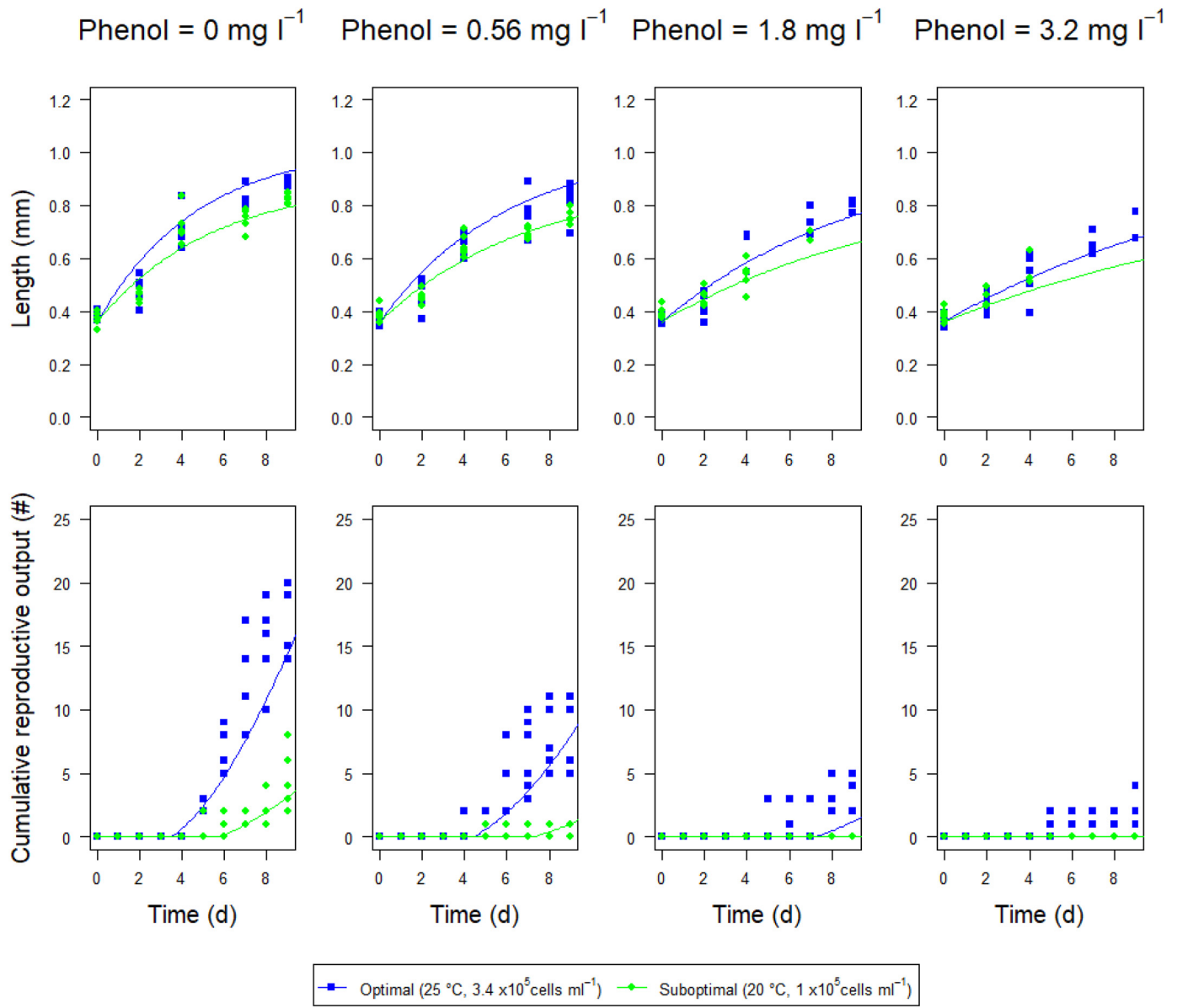

**Fig. 4.** Model predictions of combined stress effects compared to independent experimental data. Predictions (lines) of the combined effects of phenol, temperature, and food restriction on body length and reproduction in *C. dubia* using dynamic energy budget modelling. The points represent independent experimental data (i.e. model not fitted; one point per surviving organism at each time-point). The upper graphs represent growth (i.e. length) over time, and the lower graphs represent the cumulative reproductive outputs over time for four phenol concentrations and both optimal (blue squares and lines) and suboptimal (green points and lines) food levels and temperature. (For interpretation of the references to colour in this figure legend, the reader is referred to the web version of this article.)

## 4. Conclusions

This study has demonstrated that DEB theory can be used to reproduce patterns in empirical data for the joint effects of multiple environmental and chemical stressors on the life history traits of freshwater invertebrates. However, very few studies address the impact of multiple stressors for environmental conditions outside of the neutral zone (optimal range). These types of data are critical for further model development, and to challenge and refine the patterns predicted by DEB theory. Generation of more datasets similar to our study on *C. dubia* can be used to follow a pattern-oriented modelling strategy (Grimm and Railsback, 2011), as we did here.

Despite the limitations in data availability, this investigation has successfully demonstrated the plausibility of using a bioenergetics-based modelling framework to predict the effects of multiple natural and chemical stressors on life history traits. The model was corroborated by comparing its predictions for a combination of chemical, temperature, and food-related stressors with independent experimental data on *C. dubia* life history traits. To the best of our knowledge, this is the first time the effects of the interaction of three stress factors on life history traits have been successfully predicted.

Bioenergetics modelling based on DEB theory could be a powerful tool for multiple-stressor research, chemical risk assessments, and environmental management, although further work is needed to establish if DEB based bioenergetics modelling of multiple stressors also works for other freshwater species and additional compounds, and ultimately beyond freshwater invertebrates.

We suggest that existing and future multiple-stressor experiments should be reanalysed within the modelling framework outlined here to account for bioenergetics, as this reanalysis will provide a robust null model for identifying stressor interactions such as synergy and antagonism. Lastly, we expect that, data from many existing studies that concluded synergetic or antagonistic interactions could be re-analysed using the null model proposed here, and conclusions would probably need to be revised based on the improved mechanistic insight from the DEB model. Indeed, DEB based bioenergetics can already explain many interactions between stressors and reveal mechanisms behind synergism and antagonism effects.

**CRediT authorship contribution statement**

**Benoit Goussen:** Conceptualization, Methodology, Software, Formal analysis, Investigation, Data curation, Writing - original draft, Writing - review & editing. **Cecilie Rendal:** Methodology, Formal analysis. **David Sheffield:** Formal analysis. **Emma Butler:** Investigation. **Oliver R. Price:** Conceptualization. **Roman Ashauer:** Conceptualization, Methodology, Writing - original draft, Writing - review & editing.

**Declaration of competing interest**

The authors declare that they have no known competing financial interests or personal relationships that could have appeared to influence the work reported in this paper.

**Acknowledgements**

This study was funded by Unilever (MA-2014-00701). We thank our colleagues who helped with the meta-analysis literature prioritisation as well as two internal and two anonymous external reviewers that greatly helped improve this manuscript.

**Appendix A. Supplementary data**

Further details are available in the Supporting Information, including all data used in the meta-analysis, a discussion of alternative temperature relationships, case study experiments with *C. dubia* incl. statistical data analysis, details on model parameterisation, modelling of ingestion and assimilation rates and more meta-analysis results on food levels and predation threat. Supplementary data to this article can be found online at https://doi.org/10.1016/j.scitotenv.2020.141509.